\pdfoutput=1
\documentclass[12pt,a4paper]{article}

\usepackage{ifthen} 
\newboolean{pdflatex}
\setboolean{pdflatex}{true} 

\newboolean{articletitles}
\setboolean{articletitles}{true} 

\newboolean{uprightparticles}
\setboolean{uprightparticles}{false} 

\def\paperauthors{LHCb collaboration} 
\def\paperasciititle{Measurement of the branching fraction of B0 to J/psi pi0 decays} 
\def\papertitle{Measurement of the branching fraction of $\Bz \to \jpsi \piz$ decays} 
\def\paperkeywords{{High Energy Physics}, {LHCb}} 
\def\papercopyright{\the\year\ CERN for the benefit of the LHCb collaboration} 
\def\paperlicence{CC BY 4.0 licence}
\def\paperlicenceurl{https://creativecommons.org/licenses/by/4.0/}

\usepackage[top=1in, bottom=1.25in, left=1in, right=1in]{geometry}

\usepackage{microtype}
\usepackage{lineno}  
\usepackage{xspace} 
\usepackage{caption} 

\usepackage{graphicx}  
\usepackage{color}
\usepackage{colortbl}
\graphicspath{{./}} 

\usepackage{amsmath} 
\usepackage{amssymb}
\usepackage{amsfonts}
\usepackage{upgreek} 

\newcommand*\patchAmsMathEnvironmentForLineno[1]{%
\expandafter\let\csname old#1\expandafter\endcsname\csname #1\endcsname
\expandafter\let\csname oldend#1\expandafter\endcsname\csname
end#1\endcsname
 \renewenvironment{#1}%
   {\linenomath\csname old#1\endcsname}%
   {\csname oldend#1\endcsname\endlinenomath}%
}
\newcommand*\patchBothAmsMathEnvironmentsForLineno[1]{%
  \patchAmsMathEnvironmentForLineno{#1}%
  \patchAmsMathEnvironmentForLineno{#1*}%
}
\AtBeginDocument{%
\patchBothAmsMathEnvironmentsForLineno{equation}%
\patchBothAmsMathEnvironmentsForLineno{align}%
\patchBothAmsMathEnvironmentsForLineno{flalign}%
\patchBothAmsMathEnvironmentsForLineno{alignat}%
\patchBothAmsMathEnvironmentsForLineno{gather}%
\patchBothAmsMathEnvironmentsForLineno{multline}%
\patchBothAmsMathEnvironmentsForLineno{eqnarray}%
}

\usepackage{hyperxmp}

\usepackage[pdftex,
            pdfauthor={\paperauthors},
            pdftitle={\paperasciititle},
            pdfkeywords={\paperkeywords},
            pdfcopyright={Copyright (C) \papercopyright},
            pdflicenseurl={\paperlicenceurl}]{hyperref}

\usepackage[colorinlistoftodos,textsize=scriptsize]{todonotes}

\usepackage[bottom,flushmargin,hang,multiple]{footmisc}

\usepackage[all]{hypcap} 

\usepackage{xspace} 
\usepackage{upgreek}

\def\lhcb   {\mbox{LHCb}\xspace}

\def\babar  {\mbox{BaBar}\xspace}
\def\belle  {\mbox{Belle}\xspace}

\def\cleo   {\mbox{CLEO}\xspace}

\def\MagUp {\mbox{\em Mag\kern -0.05em Up}\xspace}

\ifthenelse{\boolean{uprightparticles}}%
{
 
 \def\Pgamma      {\ensuremath{\upgamma}\xspace}

 \def\Peta        {\ensuremath{\upeta}\xspace}

 \def\Pmu         {\ensuremath{\upmu}\xspace}

 \def\Ppi         {\ensuremath{\uppi}\xspace}                 
                  
 \def\Prho        {\ensuremath{\uprho}\xspace}

 \def\Ppsi        {\ensuremath{\uppsi}\xspace}

 \def\PDelta      {\ensuremath{\Delta}\xspace}                 
 \def\PXi         {\ensuremath{\Xi}\xspace}                 
 \def\PLambda     {\ensuremath{\Lambda}\xspace}                 
 \def\PSigma      {\ensuremath{\Sigma}\xspace}                 
 \def\POmega      {\ensuremath{\Omega}\xspace}                 
 \def\PUpsilon    {\ensuremath{\Upsilon}\xspace}
 \let\oldPi\Pi
 \def\PPi         {\ensuremath{\oldPi}\xspace}

 \def\PB      {\ensuremath{\mathrm{B}}\xspace}                 
                  
 \def\PD      {\ensuremath{\mathrm{D}}\xspace}

 \def\PJ      {\ensuremath{\mathrm{J}}\xspace}                 
 \def\PK      {\ensuremath{\mathrm{K}}\xspace}

 \def\Pc      {\ensuremath{\mathrm{c}}\xspace}

 \def\Pi      {\ensuremath{\mathrm{i}}\xspace}

 \def\Ps      {\ensuremath{\mathrm{s}}\xspace}

 \def\thebaroffset{0.0em}
}
{
 
 \def\Pgamma      {\ensuremath{\gamma}\xspace}

 \def\Peta        {\ensuremath{\eta}\xspace}

 \def\Pmu         {\ensuremath{\mu}\xspace}

 \def\Ppi         {\ensuremath{\pi}\xspace}                 
                  
 \def\Prho        {\ensuremath{\rho}\xspace}

 \def\Ppsi        {\ensuremath{\psi}\xspace}                 
                  
 \mathchardef\PDelta="7101
 \mathchardef\PXi="7104
 \mathchardef\PLambda="7103
 \mathchardef\PSigma="7106
 \mathchardef\POmega="710A
 \mathchardef\PUpsilon="7107
 \mathchardef\PPi="7105
                  
 \def\PB      {\ensuremath{B}\xspace}                 
                  
 \def\PD      {\ensuremath{D}\xspace}

 \def\PJ      {\ensuremath{J}\xspace}                 
 \def\PK      {\ensuremath{K}\xspace}

 \def\Pc      {\ensuremath{c}\xspace}

 \def\Pi      {\ensuremath{i}\xspace}

 \def\Ps      {\ensuremath{s}\xspace}

 \def\thebaroffset{0.18em}
}
\newcommand{\offsetoverline}[2][\thebaroffset]{\kern #1\overline{\kern -#1 #2}}%

\makeatletter
\ifcase \@ptsize \relax
  \newcommand{\miniscule}{\@setfontsize\miniscule{4}{5}}
\or
  \newcommand{\miniscule}{\@setfontsize\miniscule{5}{6}}
\or
  \newcommand{\miniscule}{\@setfontsize\miniscule{5}{6}}
\fi
\makeatother

\DeclareRobustCommand{\optbar}[1]{\shortstack{{\miniscule (\rule[.5ex]{1.25em}{.18mm})}
  \\ [-.7ex] $#1$}}

\def\mup        {{\ensuremath{\Pmu^+}}\xspace}
\def\mun        {{\ensuremath{\Pmu^-}}\xspace} 

\def\g      {{\ensuremath{\Pgamma}}\xspace}

\def\squark    {{\ensuremath{\Ps}}\xspace}

\def\cquark    {{\ensuremath{\Pc}}\xspace}

\def\pion   {{\ensuremath{\Ppi}}\xspace}
\def\piz    {{\ensuremath{\pion^0}}\xspace}
\def\pip    {{\ensuremath{\pion^+}}\xspace}
\def\pim    {{\ensuremath{\pion^-}}\xspace}

\def\rhomeson {{\ensuremath{\Prho}}\xspace}

\def\rhop     {{\ensuremath{\rhomeson^+}}\xspace}

\def\kaon    {{\ensuremath{\PK}}\xspace}

\def\KorKbar {\kern \thebaroffset\optbar{\kern -\thebaroffset \PK}{}\xspace}

\def\Kp      {{\ensuremath{\kaon^+}}\xspace}

\def\KS      {{\ensuremath{\kaon^0_{\mathrm{S}}}}\xspace}

\def\Kstarz  {{\ensuremath{\kaon^{*0}}}\xspace}

\def\Kstar   {{\ensuremath{\kaon^*}}\xspace}

\def\Kstarp  {{\ensuremath{\kaon^{*+}}}\xspace}

\newcommand{\etaz}{\ensuremath{\Peta}\xspace}
\newcommand{\etapr}{\ensuremath{\Peta^{\prime}}\xspace}

\def\D       {{\ensuremath{\PD}}\xspace}

\def\DorDbar {\kern \thebaroffset\optbar{\kern -\thebaroffset \PD}\xspace}

\def\Dp      {{\ensuremath{\D^+}}\xspace}
\def\Dm      {{\ensuremath{\D^-}}\xspace}

\def\DpDm    {\ensuremath{\Dp {\kern -0.16em \Dm}}\xspace}

\def\B       {{\ensuremath{\PB}}\xspace}
\def\Bbar    {{\ensuremath{\offsetoverline{\PB}}}\xspace}

\def\BorBbar {\kern \thebaroffset\optbar{\kern -\thebaroffset \PB}\xspace}
\def\Bz      {{\ensuremath{\B^0}}\xspace}
\def\Bzb     {{\ensuremath{\Bbar{}^0}}\xspace}
\def\Bd      {{\ensuremath{\B^0}}\xspace}

\def\BdorBdbar {\kern \thebaroffset\optbar{\kern -\thebaroffset \Bd}\xspace}
\def\Bu      {{\ensuremath{\B^+}}\xspace}

\def\Bs      {{\ensuremath{\B^0_\squark}}\xspace}

\def\BsorBsbar {\kern \thebaroffset\optbar{\kern -\thebaroffset \Bs}\xspace}
\def\Bc      {{\ensuremath{\B_\cquark^+}}\xspace}

\def\Bds     {{\ensuremath{\B_{(\squark)}^0}}\xspace}

\def\BdorBs  {\Bds}

\def\jpsi     {{\ensuremath{{\PJ\mskip -3mu/\mskip -2mu\Ppsi}}}\xspace}

\def\Y#1S{\ensuremath{\PUpsilon{(#1S)}}\xspace}

\def\LorLbar     {\kern \thebaroffset\optbar{\kern -\thebaroffset \PLambda}\xspace}

\def\BF         {{\ensuremath{\mathcal{B}}}\xspace}
\def\BR         {\BF}

\def\to                 {\ensuremath{\rightarrow}\xspace}

\def\CP                {{\ensuremath{C\!P}}\xspace}

\def\AT#1     {\ensuremath{A_{\mathrm{T}}^{#1}}\xspace}           

\def\C#1      {\ensuremath{\mathcal{C}_{#1}}\xspace}                       
\def\Cp#1     {\ensuremath{\mathcal{C}_{#1}^{'}}\xspace}                    
\def\Ceff#1   {\ensuremath{\mathcal{C}_{#1}^{\mathrm{(eff)}}}\xspace}        
\def\Cpeff#1  {\ensuremath{\mathcal{C}_{#1}^{'\mathrm{(eff)}}}\xspace}       
\def\Ope#1    {\ensuremath{\mathcal{O}_{#1}}\xspace}                       
\def\Opep#1   {\ensuremath{\mathcal{O}_{#1}^{'}}\xspace}                    

\newcommand{\nospaceunit}[1]{\ensuremath{\text{#1}}}       
\newcommand{\aunit}[1]{\ensuremath{\text{\,#1}}}       

\newcommand{\tev}{\aunit{Te\kern -0.1em V}\xspace}
\newcommand{\gev}{\aunit{Ge\kern -0.1em V}\xspace}
\newcommand{\mev}{\aunit{Me\kern -0.1em V}\xspace}
\newcommand{\kev}{\aunit{ke\kern -0.1em V}\xspace}
\newcommand{\ev}{\aunit{e\kern -0.1em V}\xspace}
 
\newcommand{\mevc}{\ensuremath{\aunit{Me\kern -0.1em V\!/}c}\xspace}
\newcommand{\gevc}{\ensuremath{\aunit{Ge\kern -0.1em V\!/}c}\xspace}
\newcommand{\mevcc}{\ensuremath{\aunit{Me\kern -0.1em V\!/}c^2}\xspace}
\newcommand{\gevcc}{\ensuremath{\aunit{Ge\kern -0.1em V\!/}c^2}\xspace}

\def\mum  {\ensuremath{\,\upmu\nospaceunit{m}}\xspace}

\def\fb   {\ensuremath{\aunit{fb}}\xspace}
\def\invfb   {\ensuremath{\fb^{-1}}\xspace}

\newcommand{\chisq}{\ensuremath{\chi^2}\xspace}

\newcommand{\chisqip}{\ensuremath{\chi^2_{\text{IP}}}\xspace}

\def\gsim{{~\raise.15em\hbox{$>$}\kern-.85em
          \lower.35em\hbox{$\sim$}~}\xspace}
\def\lsim{{~\raise.15em\hbox{$<$}\kern-.85em
          \lower.35em\hbox{$\sim$}~}\xspace}

\def\sPlot{\mbox{\em sPlot}\xspace}

\def\pt         {\ensuremath{p_{\mathrm{T}}}\xspace}

\def\ptot       {\ensuremath{p}\xspace}

\def\evtgen     {\mbox{\textsc{EvtGen}}\xspace}

\def\geant      {\mbox{\textsc{Geant4}}\xspace}

\def\photos     {\mbox{\textsc{Photos}}\xspace}

\def\pythia     {\mbox{\textsc{Pythia}}\xspace}

\def\tell1  {TELL1\xspace}
\def\ukl1   {UKL1\xspace}

\newcommand{\lhcborcid}[1]{\href{https://orcid.org/#1}{\hspace*{0.1em}\raisebox{-0.45ex}{\includegraphics[width=1em]{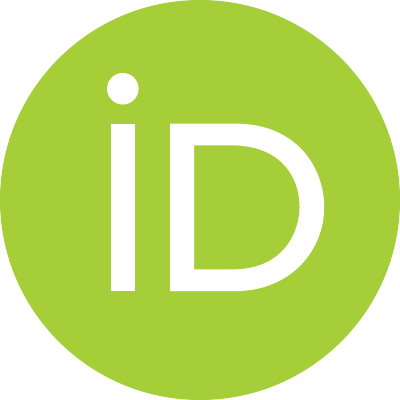}}}}

\usepackage{cite} 
\usepackage{mciteplus}

\usepackage{tikz}
\usepackage{multirow}
\usepackage{subfigure}

\begin{document}

\renewcommand{\thefootnote}{\fnsymbol{footnote}}
\setcounter{footnote}{1}


\begin{titlepage}
\pagenumbering{roman}

\vspace*{-1.5cm}
\centerline{\large EUROPEAN ORGANIZATION FOR NUCLEAR RESEARCH (CERN)}
\vspace*{1.5cm}
\noindent
\begin{tabular*}{\linewidth}{lc@{\extracolsep{\fill}}r@{\extracolsep{0pt}}}
\ifthenelse{\boolean{pdflatex}}
{\vspace*{-1.5cm}\mbox{\!\!\!\includegraphics[width=.14\textwidth]{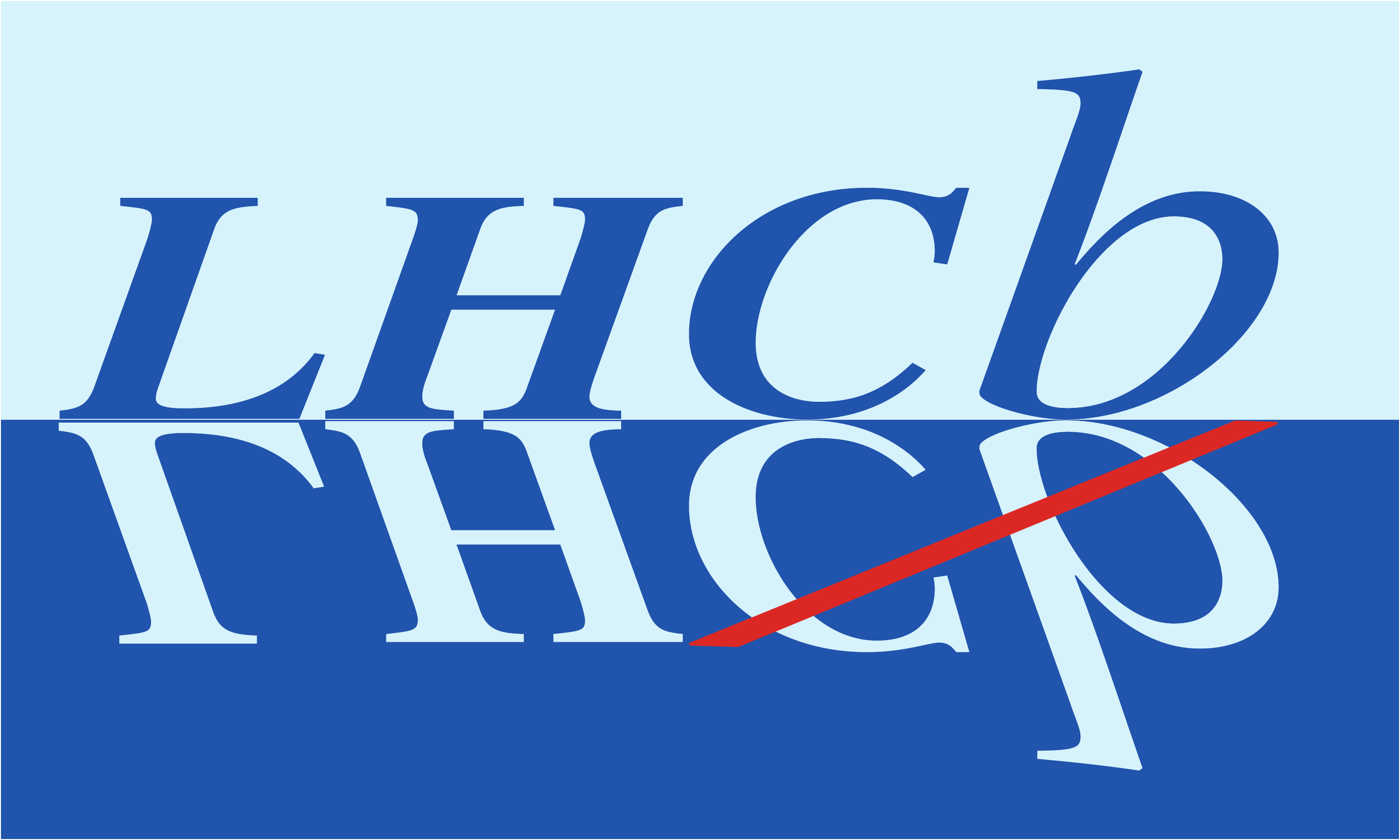}} & &}%
{\vspace*{-1.2cm}\mbox{\!\!\!\includegraphics[width=.12\textwidth]{lhcb-logo.eps}} & &}%
\\
 & & CERN-EP-2024-009 \\  
 & & LHCb-PAPER-2023-041 \\  
 & & 23 May 2024 \\ 
 
\end{tabular*}

\vspace*{4.0cm}

{\normalfont\bfseries\boldmath\huge
\begin{center}
  \papertitle 
\end{center}
}

\vspace*{1.0cm}

\begin{center}
\paperauthors\footnote{Authors are listed at the end of this paper.}
\end{center}

\vspace{\fill}

\begin{abstract}
  \noindent

  The ratio of branching fractions between \mbox{$\Bz\to\jpsi\piz$} and \mbox{$\Bu\to\jpsi\Kstarp$} decays is measured with proton-proton collision data collected by the LHCb experiment, corresponding to an integrated luminosity of $9 \invfb$.
  The measured value is

\begin{displaymath}
\frac{\mathcal{B}_{\Bz\to\jpsi\piz}}{\mathcal{B}_{\Bu\to\jpsi\Kstarp}} = (1.153 \pm  0.053 \pm 0.048 ) \times 10^{-2},
\end{displaymath}

\vspace{1.5mm}

\noindent
where the first uncertainty is statistical and the second is systematic.
The branching fraction for \mbox{$\Bz\to\jpsi\piz$} decays is determined using the branching fraction of the normalisation channel, resulting in

\begin{displaymath}
\mathcal{B}_{\Bz\to\jpsi\piz} = (1.670 \pm 0.077 \pm 0.069 \pm 0.095) \times 10^{-5},
\end{displaymath}

\vspace{1.5mm}

\noindent
where the last uncertainty corresponds to that of the external input.
This result is consistent with the current world average value and competitive with the most precise single measurement to date.
\end{abstract}

\vspace*{2.0cm}

\begin{center}
  Published in JHEP 05 (2024) 065 
\end{center}

\vspace{\fill}

{\footnotesize 
\centerline{\copyright~\papercopyright. \href{\paperlicenceurl}{\paperlicence}.}}
\vspace*{2mm}

\end{titlepage}


\newpage
\setcounter{page}{2}
\mbox{~}
%
%
%
%


\renewcommand{\thefootnote}{\arabic{footnote}}
\setcounter{footnote}{0}

\cleardoublepage


\pagestyle{plain} 
\setcounter{page}{1}
\pagenumbering{arabic}


\section{Introduction}
\label{sec:Introduction}

In the Standard Model (SM), the violation of charge conjugation-parity (\CP) symmetry in charged-current interactions between quarks is a consequence of an irreducible phase in the Cabibbo-Kobayashi-Maskawa (CKM) matrix \cite{Cabibbo:1963yz,Kobayashi:1973fv}.
The unitarity of this matrix imposes certain relations between the quark couplings, which in the \Bz sector are graphically represented by the Unitary Triangle and parameterised by three angles $\alpha$, $\beta$ and $\gamma$.
In the approximate parameterisation of the CKM matrix proposed by Wolfenstein \cite{Wolfenstein:1983yz}, the angle $\beta$ is the phase of the CKM element $V_{td}$ governing the coupling between top and down quarks, and the oscillation between \Bz and \Bzb mesons.
It thus accounts for \CP violation in the interference between direct decays of \Bz and \Bzb mesons to a \CP eigenstate and decays to the same final state after oscillation.

The angle $\beta$ is most precisely measured in the $\Bz\to\jpsi\KS$ channel\footnote{The inclusion of the charge-conjugated decays is implied throughout the paper.} where the proper-time distributions for \Bz and \Bzb decays exhibit a large asymmetry \cite{PhysRevD.23.1567}.
At tree level, these decays proceed through a Cabibbo-favoured $b \to c\overline{c} s$ quark transition, which results in a relatively large branching fraction of $\mathcal{O}(10^{-3})$.
This also facilitates the interpretation of the \CP asymmetry in terms of $\beta$ as the dominating tree-level amplitude carries no phase.
Higher-order decay topologies such as hadronic penguin diagrams can nevertheless affect the decay amplitude. By introducing a shift in the observed phase, they can mask the presence of physics beyond the SM and complicate the determination of the angle $\beta$ \cite{GROSSMAN1997241}.
While their contribution to $\Bz\to\jpsi\KS$ decays is expected to be relatively small, they might become a dominant source of uncertainty in future measurements of $\beta$ \cite{Barel:2020jvf, De_Bruyn_2023}.

Current constraints on the phase shift induced by penguin amplitudes are based on a simultaneous analysis of several \CP-violation observables for decays mediated by a $b \to c\overline{c}d$ quark transition, whose tree-level amplitude is suppressed \cite{Barel:2020jvf}.
Compared to $\Bz\to\jpsi\KS$ decays these modes are experimentally challenging; with branching fractions of $\mathcal{O}(10^{-5})$, several resonances contributing to the final state ($e.g.$ \mbox{$\Bz \to \jpsi \pip\pim$} decays \cite{LHCb-PAPER-2014-012}), photons in the final state ($e.g.$ $\Bz\to\jpsi\piz$ decays) or, in the case of \mbox{$\Bs\to\jpsi\KS$} decays \cite{LHCb-PAPER-2015-005}, a smaller production cross section for \Bs mesons compared to \Bz mesons.

The \babar and \belle collaborations reported evidence of indirect \CP-violation in \mbox{$\Bz \to \jpsi \piz$} decays \cite{Belle:2018nxw,BaBar:2008kfx}.
The reported values of the \CP observables are compatible with the results from $\Bz \to \jpsi \KS$ decays, suggesting a small contribution of loop-mediated processes to the angle $\beta$.
The world average for the branching fraction of \Bz\to\jpsi\piz  decays of $(1.66\,\pm\,0.10)\times10^{-5}$ \cite{PDG2022} is based on these two analyses.
As pointed out in Ref.~\cite{Mehraban2014}, this value could imply significant contributions from intermediate pairs of charm mesons to the decay amplitude, which makes the branching fraction an interesting probe of final-state interaction effects.
Moreover, the quark model predicts simple relations between the branching fractions of \Bz decays to $\jpsi P$ (where $P$ is a pseudo-scalar meson \piz, \etaz or \etapr) and the \etaz/\etapr mixing angles \cite{Fleischer:2011ib}, hence motivating higher precision measurements of these modes.

As a first step towards a \CP-violation analysis of $\Bz\to\jpsi\piz$ decays, this paper reports the measurement of the branching fraction through the ratio
\begin{equation}
\label{eq:brratio}
    \mathcal{R} = \frac{\BR_{\Bz\to\jpsi\piz}}{\BR_{\Bu\to\jpsi\Kstarp}},
\end{equation}

\noindent
where $\Bu\to\jpsi\Kstarp$ decays are used as the normalisation channel, and the \Kstarp meson is reconstructed via its decay to \Kp\piz.
The study is based on the dataset recorded with the LHCb detector in $pp$ collisions between 2011 and 2018, corresponding to an integrated luminosity of $9 \invfb$ collected at centre-of-mass energies of $\sqrt{s} = 7, 8$ and $13 \tev$.

\section{Detector and simulation}
\label{sec:detector}

The \lhcb detector~\cite{LHCb-DP-2012-002,LHCb-DP-2014-002} is a single-arm forward spectrometer covering the \mbox{pseudorapidity} range $2<\eta <5$, designed for the study of particles containing beauty or charm quarks. It includes a high-precision tracking system consisting of a silicon-strip vertex detector (VELO) surrounding the $pp$ interaction region, a large-area silicon-strip detector (TT) located upstream of a dipole magnet with a bending power of approximately $4{\mathrm{\,Tm}}$, and three stations of silicon-strip detectors and straw drift tubes placed downstream of the magnet. The tracking system provides a measurement of the momentum, \ptot, of charged particles with a relative uncertainty that varies from 0.5\,\% at low momentum to 1.0\,\% at 200\gevc. 
The minimum distance of a track to a primary vertex (PV), the impact parameter (IP),
is measured with a resolution of $(15+29/\pt)\mum$,
where \pt is the component of the momentum transverse to the beam, in\,\gevc.
Various charged hadrons are distinguished using information from two ring-imaging Cherenkov detectors. In addition, photons, electrons, and hadrons are identified by a calorimeter system consisting of scintillating-pad and preshower detectors, an electromagnetic and a hadronic calorimeter.
The electromagnetic calorimeter (ECAL) response is calibrated using samples of $\pi^0 \rightarrow \gamma \gamma$ decays \cite{LHCb-DP-2020-001} recorded in different detector-occupancy conditions during the 2011--2012 and 2015--2018 data-taking campaigns, respectively.
Muons are identified by a system composed of alternating layers of iron and multiwire proportional chambers.

The online event selection is performed by a trigger, which consists of a hardware stage followed by a two-level software stage \cite{LHCb-DP-2019-001}. An alignment and calibration of the detector is performed in near real-time with the results used in the software trigger~\cite{Borghi:2017hfp}. The same alignment and calibration information is propagated to the offline reconstruction, ensuring consistent information between the trigger and offline software. In this analysis, candidate events are required to pass the hardware trigger, which selects muon and dimuon candidates with high transverse momenta using information from the muon system. The first stage of the software trigger performs a partial event reconstruction and requires events to have two well-identified oppositely charged muons with an invariant mass larger than $2.7 \gevcc$. The second stage performs a 
full event reconstruction.  Events are retained for further 
processing if they contain a displaced $\jpsi \rightarrow \mu^+ \mu^-$
candidate. The decay vertex is required to be well separated 
from each reconstructed PV of the $pp$ 
interaction by requiring the distance between the PV and the $\jpsi$
decay vertex  divided by its
uncertainty to be greater than three. 

Simulated $pp$ collisions are generated using \pythia~\cite{Sjostrand:2006za,*Sjostrand:2007gs} with a specific
\lhcb configuration~\cite{LHCb-PROC-2010-056}.  Decays of hadronic
particles are described by \evtgen~\cite{Lange:2001uf}, in which
final-state radiation is generated using \photos~\cite{Golonka:2005pn}. The interaction of the generated particles with the detector, and its response, are implemented using the \geant toolkit~\cite{Allison:2006ve,*Agostinelli:2002hh} as described in Ref.~\cite{LHCb-PROC-2011-006}. 
The production of some samples is based on a computing-efficient model which re-uses the underlying event and decays the \B meson several times \cite{EPJC.78.1009}.
The resulting selection efficiencies are found to be compatible with those based on the default production model.

\section{Event selection}
\label{sec:selection}

Signal \Bz candidates are built from the combination of \jpsi and \piz mesons, reconstructed in the \mup\mun and \g\g final states, respectively.
The two muon candidates are required to have a transverse momentum larger than 500\,\mevc and to form a good vertex with a significant displacement from the PV.
Furthermore, their combination must have a mass within $100 \mevcc$ of the known \jpsi mass \cite{PDG2022}.
The photon candidates are reconstructed from isolated energy deposits in the calorimeter system with a transverse energy above 200\,\mev.
Diphoton combinations are considered only if their transverse momentum exceeds 1\,\gevc and if the associated energy deposits in the ECAL are well separated from each other. The latter requirement removes some \piz candidates at high \pt for which the mass resolution decreases.
Candidate \piz decays are retained within a wide diphoton mass window of 50-300\,\mevcc, such that the backgrounds can be studied.

Each \Bz candidate is assigned to the PV with the smallest \chisqip, defined as the smallest difference in the vertex-fit \chisq to a given PV reconstructed with and without the candidate particle being considered. A loose requirement on the \chisqip effectively
reduces combinatorial background and is complemented by the requirement that the angle between the reconstructed \Bz momentum and the direction defined by the primary and \jpsi vertices (the so-called direction angle) should be smaller than two degrees.
Furthermore, the impact parameter of the \Bz candidates should be smaller than 200\,\mum.
A kinematic vertex fit \cite{Hulsbergen:2005pu} is applied to the \Bz candidates to improve the resolution: the dimuon and diphoton masses are constrained to the known values of the \jpsi and \piz masses \cite{PDG2022}, respectively, and the \Bz candidate is assumed to have been produced at the PV.

To reject background candidates and improve the resolution on the \Bz mass, the final selection step imposes strict requirements on the particle identification (PID) of the photon, the mass of the \piz candidates, and on a multivariate classifier based on a boosted decision tree (BDT).
The classifier is trained to distinguish between simulated signal \Bz candidates and background candidates from data whose dimuon mass differs from the known \jpsi mass by more than 60\,\mevcc.
It exploits the difference between signal and background in the transverse momenta of the \Bz, \jpsi and \piz candidates, the \Bz and \jpsi IPs, the \jpsi vertex-fit \chisq and the direction angle of the \Bz candidate. Variables related to the isolation of the \jpsi vertex and to the event occupancy (such as the number of particles reconstructed around the \Bz flight direction and their momentum) are also used.
The requirements on the photon PID, diphoton mass and classifier output are chosen to maximise the product of the purity and the significance of the signal. They are determined based on the expected signal and background yields in a region around the known \Bz mass.
The BDT requirement removes more than 99.9\% of the background while retaining 25\% of the signal.
Furthermore, the diphoton mass requirement retains \piz candidates within one unit of the detector resolution, which is roughly $9\mevcc$, and reduces significantly the contamination from background events.

The decay mode $\Bu\to\jpsi\Kstarp(\to\Kp\piz)$ is used for the normalisation. The \jpsi and \piz candidates are selected with identical criteria as for the signal mode, while the \Kp meson is required to have a \pt greater than 250\,\mevc, pass loose PID requirements, and be associated to a track that is significantly displaced from the PV. The mass of the \Kp\piz combination must lie within $100 \mevcc$ of the known \Kstarp mass \cite{PDG2022}.
Finally, \Bu and \Bz candidates share the same selection criteria for the classifier output.

\section{Yields of signal and normalisation decays}
\label{sec:fitmodel}

The diphoton mass distributions in the signal and normalisation data samples are first studied to constrain the yield of some background contributions, as described in Sec.~\ref{sec:fitmodel_pi0}.
These constraints are then used to determine the yield of signal decays based on a fit to the \Bz mass distribution in data, as presented in Sec.~\ref{sec:fitmodel_jpsipi0}.
Finally, the yield of \Bu decays is reported in Sec.~\ref{sec:fitmodel_jpsikst}.

\subsection{Study of the diphoton mass distribution}
\label{sec:fitmodel_pi0}

The mass distribution of \piz candidates reflects the presence of partially reconstructed \mbox{$\KS\to\piz\piz$} decays where one pion is missed. In the signal data sample, these partially reconstructed decays represent a significant source of background, especially from processes such as $\Bz\to\jpsi\KS$ decays.
The sensitivity to the \KS contamination stems from the significant flight distance of \KS mesons in the LHCb tracking system and the hypothesis used at the reconstruction level that photons originate from the interaction region. Neutral pions from \KS decays are thus reconstructed with a wrong production vertex and on average, at a lower mass than \piz mesons from signal decays.
This signature is exploited to constrain the \KS contamination in the data sample by a fit to the diphoton mass distribution.
Furthermore, these false \piz candidates representing combinatorial background are broadly distributed in the diphoton mass, which can also be exploited to assess the contamination of that background.

The yields of \KS and false \piz background contributions in the signal data sample are first determined over the full diphoton mass range and then extrapolated to the narrow region around the known \piz mass, that corresponds to the final selection requirement.
The yields are determined from an extended unbinned maximum-likelihood fit to the diphoton mass distribution, which includes events with a genuine \piz meson produced near the interaction region, thus exhibiting a correctly reconstructed mass around $135 \mevcc$.
The mass shapes of the three components are based on simulation.
The main contributions to the genuine \piz and \KS components are $\Bz\to\jpsi\piz$ and $\Bz\to\jpsi\KS$ decays, respectively. Simulated samples corresponding to these decays are thus used to parameterise the shapes of the two components.
The sum of two Crystal Ball functions \cite{Gaiser:1982yw} that share the same mean and width parameters (later referred to as a double-sided Crystal Ball function) is used for the shape of the genuine \piz component while a Crystal Ball function with a different width for masses below and above the mean represents the \KS component.
The false \piz component is modelled by a power-law function using samples of simulated $\B_{u,d,s}\to\jpsi X$ decays, where $X$ refers to different sets of final-state particles produced in the known \B decay modes with branching fractions larger than roughly $10^{-5}$.

All shape parameters are fixed to the values found in the fits to simulated distributions, except for the mean and widths of the Crystal Ball functions that are corrected using a similar fit to the normalisation data sample, where \KS backgrounds are negligible and these corrections are well measured.

\begin{figure}[t!]
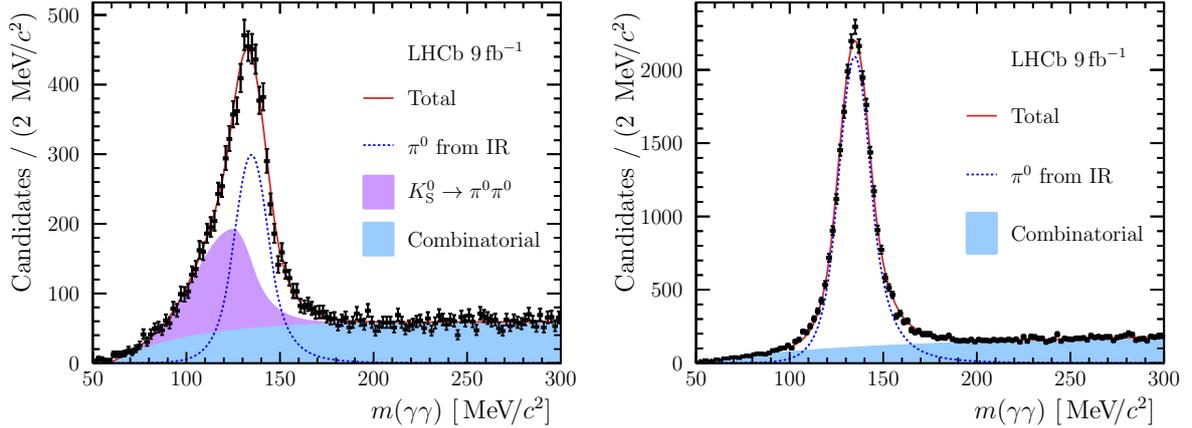

\begin{center}
\scalebox{0.38}{\input{fig1a.tex}}
\scalebox{0.38}{\input{fig1b.tex}}
\caption{\small Mass distribution of \piz candidates shown with the fit projection for the (left) signal and (right) normalisation modes. The dashed line and coloured regions represent the \piz produced at the interaction region (IR) and the other fit components, respectively.}
\label{fig:fit_model_pi0}
\end{center}
\end{figure}

The diphoton mass distributions for the signal and normalisation data samples are shown in Fig.~\ref{fig:fit_model_pi0}, without the diphoton mass requirement, together with the fit results.
In the narrow region around the known \piz mass, $832\pm82$ and $449\pm27$ \KS and false \piz candidates are found, respectively. The uncertainties account for the choice of parameterisation of the fit components and the precision of the mean and width corrections.

\subsection{Signal yield}
\label{sec:fitmodel_jpsipi0}

The number of signal decays is determined using an extended unbinned maximum-likelihood fit to the measured \jpsi\piz mass distribution.
The signal component of the fit is parameterised by the sum of two Crystal Ball functions that share the same mean.
The shape parameters are fixed to the values obtained from the simulation, except the widths $\sigma_{\rm L,R}$ on the left and right side of the peak, which are allowed to deviate from their simulated values by a scale factor $R_{\sigma}=\sigma_{\textrm{data}}/\sigma_{\textrm{sim}}$ that is free to vary in the fit.

The fit model accounts for two main sources of background.
First, several fit components are associated to partially reconstructed $\B\to\jpsi\piz\pi$ decays where the second pion, neutral or charged, is missed.
Each corresponding mass shape is parameterised by a double-sided Crystal Ball function using simulation.
As done for the signal shape, the simulated width parameters are corrected by $R_{\sigma}$.
Based on their known dependence on the mass resolution, the mean and tail parameters are also corrected.
While the yields of $\B_{u,c}\to\jpsi\rhop(\to\pip\piz)$ decays are free to vary in the fit, the yield of \mbox{$\Bz\to\jpsi\KS$} decays is fixed to the \KS yield found in the diphoton mass fit.
A small contribution from \mbox{$\B\to\jpsi\Kstar(\to\KS\pion)$} decays (with $\KS\to\piz\piz$ and one of the two \piz is used) is expected but neglected in the fit as its impact on the signal yield is negligible (see Sec.~\ref{sec:syst_jpsipi0}).
Other $\B\to\jpsi X$ backgrounds where $X=(\phi,\eta,\omega)$ decays to \pip\pim\piz, or $X=\omega$ (\Kstar) and decays to \g\piz ($K\piz$) were investigated and found to be negligible.

\begin{figure}[t!]
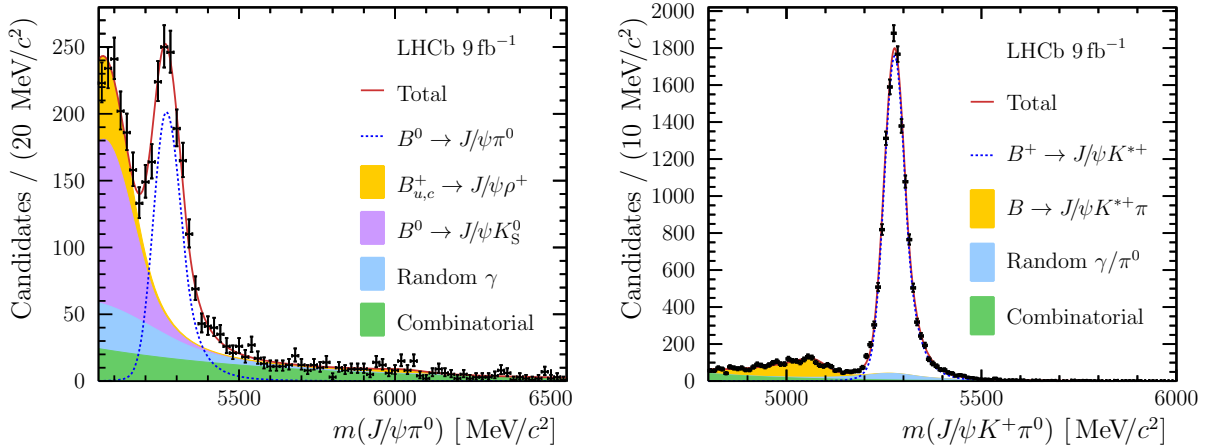

\begin{center}
\scalebox{0.38}{\input{fig2a.tex}}
\scalebox{0.38}{\input{fig2b.tex}}
\caption{\small Mass distribution of \B candidates shown with the fit projection for (left) signal and (right) normalisation modes. The dashed line and coloured regions represent $\Bz\to\jpsi\piz$ or $\Bu\to\jpsi\Kstarp$ decays and different sources of background, respectively.}
\label{fig:fit_model_jpsipi0}
\end{center}
\end{figure}

As a second source of background, one or more final-state particles in a $\it{b}$-hadron decay can be mistakenly replaced by a particle from the underlying event. In that case, the mass of the corresponding \jpsi\piz candidates will peak close to the signal region.
Due to the vertex requirements applied to charged particles, the probability of an occurrence is only significant when the final-state particles are photons: either one decay photon is replaced which results in a false \piz candidate, or two decay photons are replaced by a false or genuine \piz meson.
The \jpsi\piz mass shape of false \piz contributions
from signal and partially reconstructed decays (later referred to as Random $\gamma$ in Fig.\,\ref{fig:fit_model_jpsipi0} (left)) are determined using \piz candidates with masses above 200\,\mevcc. A parameterisation based on the sum of an exponential function and a Gaussian function with exponential tails is used.
Their contamination in the narrow region around the known \piz mass is fixed to the yield of false \piz candidates found in the diphoton mass fit.
Although they do not involve photons from the underlying event, false \piz candidates formed in $e.g.$ $\Bz\to\jpsi\KS(\to\piz\piz)$ decays by combining photons from the two different \piz decays are implicitly accounted for in this approach.
Finally, background contributions from genuine \piz mesons (later referred to as Combinatorial in Fig.\,\ref{fig:fit_model_jpsipi0} (left)) are more broadly distributed than the false \piz contributions and are therefore modelled differently, using an exponential function whose parameter $\alpha_{\rm B}$ is free to vary in the fit.

A yield of $1232\pm55$ $\Bz\to\jpsi\piz$ decays is measured in the data sample.
The mass distribution of \jpsi\piz candidates and the fit results are shown in Fig.~\ref{fig:fit_model_jpsipi0} (left) where the upper bound of the mass range is lower than that of the fit by $400 \mevcc$ to emphasise the \Bz peak.
All fit parameters are listed in Table~\ref{tab:fit_model_jpsipi0}.
The yields of partially reconstructed $\Bu\to\jpsi\rhop$ and $\Bc\to\jpsi\rhop$ backgrounds determined by the signal fit and their expected values based on the full reconstruction of the decays (as presented in Sec.~\ref{sec:checks_bkg}) are consistent within 0.5 and 0.2 standard deviations, respectively.

\begin{table}[tb]
\renewcommand{\arraystretch}{1.2}
\caption{Results of the fit to the \jpsi\piz mass: the yields of signal and background decays, the yield of genuine \piz combinatorial candidates ($N_{\rm B}$), the ratio between the signal width in data and in simulation ($R_{\sigma}$), and the shape parameter of the combinatorial background ($\alpha_{\rm B}$).}
\begin{center}
\begin{tabular}{cc} 
    \hline
    Parameter & $\phantom{0-.00000}\text{Value}\phantom{\pm 000.}$ \\
    \hline
    $N(\Bz\to\jpsi\piz)$  & $\phantom{0-.}1232   \pm 55\phantom{0.}$  \\
    $N(\Bu\to\jpsi\rhop)$ & $\phantom{0-0.}307   \pm 49\phantom{0.}$  \\
    $N(\Bc\to\jpsi\rhop)$ & $\phantom{0-00.}75   \pm 30\phantom{0.}$ \\
    $N_{\rm B}$           & $\phantom{0-0.}783   \pm 91\phantom{0.}$ \\
    $R_{\sigma}$          & $\phantom{0-0}0.88   \pm 0.04$  \\
    $\alpha_{\rm B}\times10^{3}$ $(c^{2}/\mev)$ & $\phantom{00}-1.56 \pm 0.16$ \\
    \hline
\end{tabular}
\end{center}
\label{tab:fit_model_jpsipi0}
\end{table}

\subsection{Normalisation decay yield}
\label{sec:fitmodel_jpsikst}

The yield of normalisation decays is determined using an extended binned maximum-likelihood fit to the measured \jpsi\Kp\piz mass distribution.
The signal component of the fit is parameterised with a double-sided Crystal Ball function.
Shape parameters are taken from simulation while the mean \jpsi\Kstarp mass and width parameters are allowed to deviate from their simulated values by an offset $D_{\mu,\Kstar} = \mu_{\textrm{data},\Kstar} - \mu_{\textrm{sim},\Kstar}$ and a scale factor $R_{\sigma,\Kstar}$, respectively.

Partially reconstructed background contributions are dominated by $\B\to\jpsi\Kstarp\pi$ decays where the pion is missed.
 Although several kaon resonances contribute to the three-hadron final state, the corresponding mass shape is determined using simulation samples of $\Bu\to\jpsi K_{1}(1270)^{+}$ decays (where the $K_{1}(1270)^{+}$ resonance decays to \Kstarp\piz) and parameterised by a double-sided Crystal Ball function.
The \jpsi\Kp\piz mass distributions for kaon resonances heavier than $K_{1}(1270)^{+}$ mesons are expected to be shifted to lower \jpsi\Kp\piz masses due to the larger momentum of the missed pion.
The mean parameter is therefore free to vary in the fit to account for the presence of heavier resonances.
The width, however, is corrected using the same scale factor $R_{\sigma,\Kstar}$ as used for the normalisation mass shape.
Other parameters are fixed to the values extracted from simulation. 

Background \Bu candidates formed when mistakenly associating one or two photons from the underlying event to the final-state particles produced in a $\it{b}$-hadron decay are also present in the normalisation data sample and mainly stem from $\Bu\to\jpsi\Kstarp$ decays due to the large branching fraction and high selection efficiency.
As a result, they can be simply parameterised using the corresponding simulation samples instead of the data-driven approach followed for the signal model (the two approaches are compared in Sec.~\ref{sec:checks_assumptions}).
The components associated with one- and two-photon background contributions are modelled as a double-sided Crystal Ball function and a Gaussian function with exponential tails, respectively.
In both cases, the mean and width parameters are subject to the corrections $D_{\mu,\Kstar}$ and $R_{\sigma,\Kstar}$ applied to the normalisation shape parameters.
The two background yields are expressed as the product of the yield of normalisation decays times their relative contributions expected from simulation.
These contributions are corrected for the slightly larger photon occupancy in data and for isospin-conjugated $\Bz\to\jpsi\Kstarz(\to\Kp\pim)$ decays (where the missed charged pion is replaced by a \piz  candidate from the underlying event) which contribute to the two-photon background only.
To a lesser extent, partially reconstructed $\B\to\jpsi\Kstarp\pi$ decays can also be misreconstructed by exchanging the decay photons with photons from the event. To ensure that this component is accounted for, the reconstructed photons in the $\Bu\to\jpsi K_{1}(1270)^{+}$ sample are allowed to originate from the underlying event.
The yield of that component is not corrected for the slightly larger photon occupancy in data, however, the impact of this correction on the normalisation yield is negligible.
Finally, combinatorial background contributions caused by the wrong association of final-state particles with a muon or a kaon from the event are modelled using an exponential function, with its slope parameter $\alpha_{\textrm{B},\Kstar}$ freely varying in the fit.

\clearpage

Other sources of peaking backgrounds were investigated and found to be negligible given their lower branching fraction and selection efficiency, such as $\Bu\to\jpsi\rhop(\to\pip\piz)$ decays, where the charged pion is misidentified as a kaon, and $\Bu\to\chi_{\rm c1}(\to\jpsi\gamma)\Kp$ decays,
where the photon is associated to a photon from the event.

In the data sample, the yield of $\Bu\to\jpsi\Kstarp$ decays is measured to be \mbox{$13052\pm115$}.
The mass distribution of \jpsi\Kp\piz candidates and the fit results are shown in Fig.~\ref{fig:fit_model_jpsipi0} (right) while the fitted parameters are listed in Table~\ref{tab:fit_model_jpsikst}.
The simulation effectively reproduces the \Bu peak position for normalisation decays which justifies that this parameter is fixed in the fit to the signal data sample.
Contrary to the signal mode where the \Bz mass resolution was better in data than in simulation, the scale factor is slightly above one for the normalisation mode.
Finally, the shift in mass for the shape of partially reconstructed $\B\to\jpsi\Kstarp\pi$ decays is negative, as expected from the contributions of kaon resonances heavier than the $K_{1}(1270)^{+}$ meson.

\begin{table}[tb]
\renewcommand{\arraystretch}{1.2}
\caption{Results of the fit to the \jpsi\Kp\piz mass: the yields of normalisation and partially reconstructed background decays, the yield of combinatorial candidates ($N_{\textrm{B},\Kstar}$), the ratio between the normalisation width in data and in simulation ($R_{\sigma,\Kstar}$), the difference between the normalisation mean in data and in simulation ($D_{\mu,\Kstar}$), the difference $D_{\mu,\Kstar\pi}$ for partially reconstructed decays and the shape parameter of the combinatorial background ($\alpha_{\textrm{B},\Kstar}$).}
\begin{center}
\begin{tabular}{cc}
    \hline
    Parameter & $\phantom{0-.00000}\text{Value}\phantom{\pm 000.}$ \\
    \hline
    $N(\Bu\to\jpsi\Kstarp)$      & $\phantom{0-.}13052 \pm 115\phantom{.}$ \\
    $N(B\to\jpsi\Kstarp\pi)$     & $\phantom{0-0.}1998  \pm 79\phantom{0.}$\\
    $N_{\textrm{B},\Kstar}$      & $\phantom{0-0.}1039  \pm 90\phantom{0.}$ \\
    $R_{\sigma,\Kstar}$          & $\phantom{0-00}1.05  \pm 0.01$ \\
    $D_{\mu,\Kstar}$ (\mevcc)    & $\phantom{0000}-0.3 \pm 0.3\phantom{0}$ \\
    $D_{\mu,\Kstar\pi}$ (\mevcc)                            & $\phantom{0000.}-17 \pm 2\phantom{00.}$ \\
    $\alpha_{\textrm{B},\Kstar}\times10^{3}$ $(c^{2}/\mev)$ & $\phantom{0000}-3.9 \pm 0.2\phantom{0}$ \\
    \hline
\end{tabular}
\end{center}
\label{tab:fit_model_jpsikst}
\end{table}

\section{Branching fraction ratio result}
\label{sec:result}

The ratio of the branching fractions between the signal and normalisation decays defined in Eq.~\ref{eq:brratio} is calculated as

\begin{equation}
\label{eq:brratio2}
    \mathcal{R} = \frac{N_{\Bz\to\jpsi\piz}}{N_{\Bu\to\jpsi\Kstarp}} \times \frac{\epsilon_{\Bu\to\jpsi\Kstarp}}{\epsilon_{\Bz\to\jpsi\piz}} \times \BR_{\Kstarp\to\Kp\piz},
\end{equation}

\noindent
where $N_{\Bz\to\jpsi\piz}$ and $N_{\Bu\to\jpsi\Kstarp}$ are the yields reported in Sec.~\ref{sec:fitmodel}, $\epsilon_{\Bz\to\jpsi\piz}$ and $\epsilon_{\Bu\to\jpsi\Kstarp}$ the efficiencies for the selection requirements applied to the corresponding candidates, and $\BR_{\Kstarp\to\Kp\piz}=1/3$ \cite{PDG2022}.
The efficiencies account for the loss of \B-meson candidates due to the detector acceptance, the reconstruction of the final-state particles and the selection requirements detailed in Sec.~\ref{sec:selection}.

The efficiencies that are not associated with PID requirements are estimated using simulated samples without any particular weighting of the events, as the known discrepancies between data and simulation are small.
Furthermore, the bias should be comparable in the two modes and cancel in the efficiency ratio.
As assessed in detail in Sec.~\ref{sec:syst_eff}, the systematic uncertainty assigned to this assumption is relatively small compared to the statistical uncertainty affecting the signal yield.

Photon PID efficiencies are also estimated using simulation.
Due to a strong dependence of the PID performance on the particle multiplicity, a weight is applied to each simulated event to improve the description of the photon cluster isolation in the ECAL, the photon kinematics and the number of reconstructed tracks, and eventually of the PID response.
The weights are determined using the normalisation data sample from which background contributions are statistically removed by the \sPlot method \cite{Pivk:2004ty}, and are applied to both the normalisation and signal simulated samples.
The efficiencies are then calculated from the sums of weights over the events passing and failing the photon PID requirement.
In the selection of the normalisation candidates, the efficiency for the charged kaon PID requirement is determined using calibration samples collected from charm decays \cite{Anderlini:2202412}.

Finally, as the reconstructed \piz mass in 2011--2012 data samples is above the known \piz mass by 1--2\%, the simulation is corrected accordingly before estimating the efficiency of the selection requirements.

An efficiency ratio between the signal and normalisation decays of $2.73\pm0.05$ is obtained, the normalisation mode efficiency being smaller mostly because of the kaon reconstruction requirement.
Using Eq.~\ref{eq:brratio2}, the obtained ratio of branching fractions is

\begin{equation*}
    \mathcal{R} = (1.153 \pm  0.053 ) \times 10^{-2},
\label{eq:result1}
\end{equation*}
\noindent
where the uncertainty is statistical.

\section{Systematic uncertainties}
\label{sec:syst}

\subsection{Signal fit model}
\label{sec:syst_jpsipi0}

The dominant sources of uncertainty are associated to the choice of parameterisation for the mass shapes in the signal fit, the constraints on the \KS and false \piz background contributions, the assumption of negligible $B\to\jpsi\KS\pi$ contributions, and the choice to set the mean parameter of the signal shape to its simulated value ($i.e.$ $D_{\mu}=0$).

A different parameterisation is used for each fit component and the modified model is fitted to the data sample.
The variation in signal yield is taken as a systematic uncertainty, except when more than one alternative parameterisation is considered, in which case the average variation or the standard deviation is considered.

The signal shape is replaced by a double-sided Novosibirsk function \cite{Belle:1999bhb} while random variations of the default parameters describing partially reconstructed backgrounds are generated using the covariance matrices. A hundred sets of new parameters are obtained for each background and used in the fit, and the standard deviation of the signal yield is taken as a systematic uncertainty.
In the default model, combinatorial backgrounds associated to photons from the event are modelled using the high-mass region of the diphoton mass distribution that extends from 200--300\,\mevcc.
This interval is divided in four regions to determine new values of the shape parameters, keeping the same function. The average variation of the signal yield for the four parameterisations is taken as a systematic uncertainty.
An alternative description for the other combinatorial background component is obtained by combining \jpsi and \piz candidates from different events.
The resulting \Bz candidates are passed through the full chain of offline selection requirements and their mass distribution is parameterised by the sum of a Crystal Ball function and an exponential function with a slope free to vary in the fit.
The fit results using these models deviate from the default result by less than 1\%, except in the case of partially reconstructed \mbox{$\Bz\to\jpsi\KS$} background decays, where a change in signal yield of 1.5\% is observed.

In the default model, the yields of \KS and false \piz backgrounds are fixed to the values obtained in the study of the diphoton mass distribution.
The default fit is repeated after changing the yields by their uncertainty, as reported in Sec.~\ref{sec:fitmodel_pi0}.
The resulting changes in signal yield lie between 0.6--0.9\% and are considered as a systematic uncertainty.

Previously neglected $B\to\jpsi\KS\pi$ decays are incorporated in the fit model by setting their contribution relative to $\Bz\to\jpsi\KS$ decays to its expected value. As these events are distributed well below the signal peak, the resulting change in signal yield is negligible.

Finally, the 1--2\% difference in the ECAL calibration between the 2011--2012 and 2015--2018 data-taking campaigns is accounted for by leaving the signal mean parameter $D_{\mu}$ free to vary in the fit.
A change in signal of 1.5\% is observed while the parameter $D_{\mu}$ takes the value $(3.8\pm2.2) \mevcc$.

In summary, a total uncertainty associated to the choice of signal fit model of 2.6\% is found, dominated by the parameterisation of the \KS background shape and the ECAL energy scale.

\subsection{Normalisation fit model}
\label{sec:syst_jpsikst}

The strategy to assess the uncertainties on the yield of normalisation decays is similar to that of the signal yield.
The shape of the normalisation component is replaced by a Bukin function \cite{Bukin}.
For the background components related to underlying-event photons, combinatorial candidates and partially reconstructed decays, the default functions are replaced by the sum of two Gaussian functions, a Chebyshev polynomial of second degree \cite{Chebyshev} and an Argus function \cite{ARGUS:1990hfq} convoluted with a Gaussian function, respectively.
The largest variations in the normalisation yield are found for the modified model of the normalisation and partially reconstructed decays, which are both at the level of 0.3--0.4\%.

The correction to the relative yield of one- and two-photon backgrounds with respect to that of the normalisation yield is changed according to its uncertainty, as determined in a study of $\Bu\to\jpsi\Kp$ decays that contaminate the data sample when the kaon is mistakenly associated with two photons from the event.
A systematic uncertainty of 0.3\% is assigned.

The different ECAL calibration in the 2011--2012 and 2015--2018 data samples observed at the \piz mass level (1--2\%) is reflected in the \Bu mass (0.06\%) despite the \piz mass constraint imposed by the kinematic fit of the decays.
This small difference in the \Bu mass is ignored in the default model, which determines the shape of the normalisation channel from simulated samples that have the same calibration.
The associated uncertainty is determined by correcting the simulated \Bu mass of the 2011--2012 and 2015--2018 samples to match the mass measured in the two different data sets, deriving new parameters for the normalisation shape and fitting the new model to the combined data set.
The relative change in yield of 0.2\% is taken as a systematic uncertainty, which brings the total uncertainty associated with the choice of the normalisation model to 0.6\%.

\subsection{Selection efficiencies}
\label{sec:syst_eff}

Given the similar topology and selection of signal and normalisation decays, the ratio of efficiency for several requirements should be close to one.
This is assessed in detail for the BDT classifier, trigger and photon PID requirements by comparing the corresponding efficiencies in data and in simulation.

The multivariate classifier is constructed to have a similar response to signal and normalisation decays, but also to their isospin-conjugated decays $\Bu\to\jpsi\pip$ and \mbox{$\Bz\to\jpsi\Kstarz$}, respectively.
These two modes are selected with relatively high efficiency and are less prone to backgrounds due to the absence of a neutral pion in the final state.
They are therefore suitable to measure precisely the efficiency of the classifier output requirement and derive a possible correction to the efficiency ratio appearing in Eq.~\ref{eq:brratio2}.
That correction is measured with a precision of 1.7\%, limited by the size of the $\Bu\to\jpsi\pip$ data sample.
Given that the difference between the measured and simulated efficiency ratios is less than that precision, no correction is applied and that precision is taken as a systematic uncertainty.

The uncertainty associated with the trigger requirements is determined similarly using samples of $\Bu\to\jpsi\Kp$ events that fulfil any trigger requirements, with the \jpsi signals excluded~\cite{LHCb-DP-2012-004}.
Each sample is divided into two smaller samples, which are weighted to reproduce the muon kinematics in the signal and normalisation modes, respectively.
The obtained efficiency ratio in data and simulation shows an excellent compatibility of the order of 0.3\% which is considered as a systematic uncertainty.

The modelling of the photon PID distributions is studied using background-subtracted samples of normalisation decays.
After dividing the sample into two halves, one of the two samples is further weighted such that the photon kinematic and isolation variables match the ones for signal decays.
The efficiency is then determined for a wide range of photon PID requirements in each of the two samples and the obtained values compared to the simulation.
The difference in the efficiency ratios between data and simulation at the chosen PID requirement is negligible, however, the largest difference of 0.8\% is retained as a conservative estimate of the systematic uncertainty.

Uncertainties specific to the selection of $\Bu\to\jpsi\Kstarp$ candidates pertain to the kaon reconstruction and PID requirement, as well as the \Kstarp mass requirement.
The former is evaluated based on dedicated track reconstruction studies performed with \mbox{$\jpsi\to\mup\mun$} decays whose results are extrapolated to the case of kaons, and gives an uncertainty of 1.5\% \cite{collaboration_2015}.
The PID requirement efficiency is derived using calibration samples from which three-dimensional efficiency tables in bins of kaon \pt, $\eta$, and the number of reconstructed tracks in the event are produced. The influence of the choice of binning on the efficiency is found to be below 0.3\% and is taken as a systematic uncertainty.
The influence of the size of the calibration samples is below one part per thousand and is therefore ignored.
Finally, the modelling of the \Kstarp mass shape is studied by enlarging the selection mass window of \Kp\piz candidates (from $\pm100$ to $\pm300 \mevcc$ around the known \Kstarp mass) for \jpsi\Kp\piz candidates in a narrow region around the \Bu mass peak.
This pure sample of \Kstarp mesons allows a precise determination of the efficiency in data, compatible with the simulated value at the level of 1.1\%, which is taken as a systematic uncertainty.

The size of the simulation sample for signal and normalisation decays enters the total systematic uncertainty through the statistical precision of the estimated selection efficiencies, which is evaluated to 1.6\%.

\subsection{Summary of uncertainties}

The sources of systematic uncertainties affecting the ratio of branching fractions are summarised in Table~\ref{tab:syst_summary}.
Statistical and systematic uncertainties are found to be of the same order and are added in quadrature.

\begin{table}[tb]
\renewcommand{\arraystretch}{1.1}
\caption{Summary of uncertainties on the ratio of branching fractions for \mbox{$\Bz\to\jpsi\piz$} and \mbox{$\Bu\to\jpsi\Kstarp$} decays. Systematic uncertainties  of individual sources are added in quadrature.}
\begin{center}\begin{tabular}{lc}
   \hline
    Source & Uncertainty (\%)\\
   \hline
    Signal model & 2.6 \\
    Normalisation model & 0.6 \\
    Simulated sample size & 1.6 \\
    BDT classifier & 1.7 \\
    Trigger & 0.3 \\
    Photon PID & 0.8 \\
    Kaon reconstruction & 1.5 \\
    \Kstarp mass & 1.1 \\
    \hline
    Systematic uncertainty & 4.1 \\
    Statistical uncertainty & 4.6\\
    Total uncertainty & 6.2\\
   \hline
\end{tabular}\end{center}
\label{tab:syst_summary}
\end{table}

\section{Cross-checks}
\label{sec:checks}

\subsection{Study of background modes}
\label{sec:checks_bkg}

The number of $307\pm49$ partially reconstructed \Bu decays to the \jpsi\rhop final state found in the signal fit (Table~\ref{tab:fit_model_jpsipi0}) is compatible with the expected yield of $339\pm19$ based on a reconstruction of this mode that includes the charged pion.
Similarly, the contribution of $75\pm30$ $\Bc\to\jpsi\rhop$ decays to the signal data sample is also compatible with the value of $62\pm4$ obtained when fully reconstructing these decays.

In the study of these backgrounds, the selection of \rhop candidates is similar to those of the \Kstarp in normalisation decays after exchanging the kaon for a pion.
Moreover, the mass of \pip\piz candidates is required to be within $200 \mevcc$ of the known \rhop mass \cite{PDG2022}.
The \Bu candidates are selected with identical criteria on the BDT classifier, \piz mass and photon PID as for the signal mode. In the case of \Bc candidates, these selection requirements are relaxed  to accommodate the small production of \Bc mesons.

The determination of the yield of \Bu decays accounts for partially reconstructed $\Bu\to\jpsi\Kstarp$ background contributions where \Kstarp mesons decay to $\KS(\to\piz\piz)\pip$ and one \piz meson is missed.
Like in the signal mode, decays with a \KS meson in the final state and random photon backgrounds are constrained using a fit to the diphoton mass distribution.
Backgrounds from the mis-identification of the kaon in normalisation decays are suppressed by a dedicated PID requirement imposed on the pion.
In the \Bc sample, peaking backgrounds are ignored and only combinatorial backgrounds are modelled in the fit.

\begin{figure}[t!]
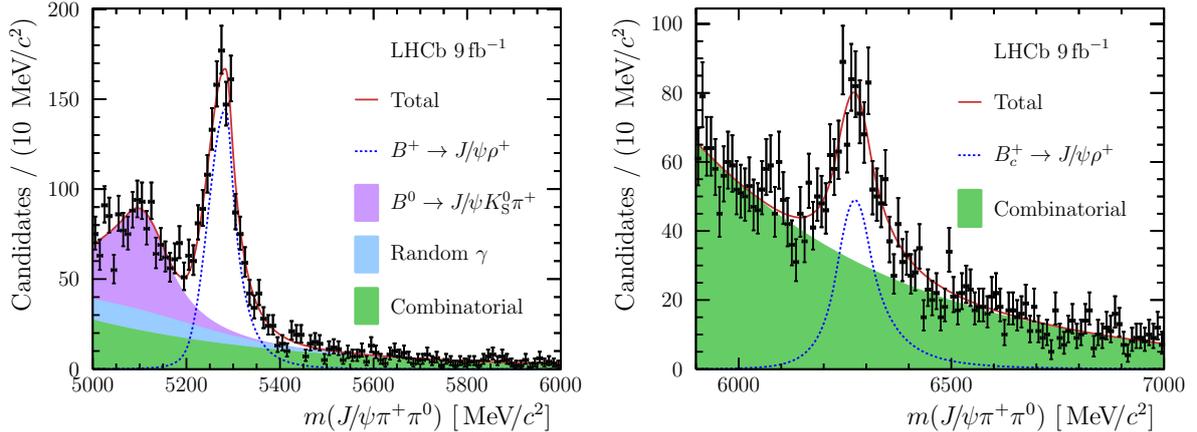

\begin{center}
\scalebox{0.38}{\input{fig3a.tex}}
\scalebox{0.38}{\input{fig3b.tex}}
\caption{\small Mass distribution of \jpsi\rhop candidates shown with the fit projection in the (left) \Bu and (right) \Bc signal region. In the latter case, the selection requirements are relaxed to accommodate the small production of \Bc mesons. The dashed line and coloured regions represent \Bu or \Bc decays to the \jpsi\rhop final-state and different sources of background, respectively.}
\label{fig:fit_model_jpsirho}
\end{center}
\end{figure}

Results of the fits to the \Bu and \Bc samples are shown in Fig.~\ref{fig:fit_model_jpsirho}.
Given that the \Bc mode has never been observed, the detailed analysis of these decays is the subject of a separate publication \cite{LHCb-PAPER-2023-046}.

\subsection{Study of isospin-conjugated modes}
\label{sec:checks_iso}

Decays of $\Bu\to\jpsi\pip$ and $\Bz\to\jpsi\Kstarz$ are used to assess the systematic uncertainty associated to the modelling of the BDT classifier response.
They can also be used to verify that the total efficiencies are well estimated in simulation by measuring the ratio of the branching fractions.

In this study, the \pt requirement for the \pip meson is identical to the one for the \piz mesons in signal and normalisation decays.
In addition to the previous selection requirements placed on the \Kp and \Kstar candidates, stringent PID requirements are imposed to reduce the contamination from the main peaking backgrounds.
Accordingly, the mass model for \jpsi\pip background candidates accounts only for mis-identified $\Bu\to\jpsi\Kp$ decays which still pass the PID selections given a much larger branching fraction, and a combinatorial component.
The \jpsi\Kstarz model includes partially reconstructed decays with one missed pion and a combinatorial component.
Results of the fits to the \Bu and \Bz samples are shown in Fig.~\ref{fig:fit_model_iso}.
The obtained branching fraction ratio is compatible with the world average at the level of 1.3 standard deviations.

\begin{figure}[t!]
\begin{center}
\scalebox{0.38}{\input{fig4a.tex}}
\scalebox{0.38}{\input{fig4b.tex}}
\caption{\small Mass distribution of (left) \jpsi\pip and (right) \jpsi\Kstarz candidates shown with the fit projection. The dashed lines and coloured regions represent $\Bu\to\jpsi\pip$ or $\BdorBs\to\jpsi\Kstarz$ decays and different sources of background, respectively.}
\label{fig:fit_model_iso}
\end{center}
\end{figure}

\subsection{Assumptions used in the signal and normalisation fits}
\label{sec:checks_assumptions}

Partially reconstructed $\Bz\to\jpsi\KS$ and $\Bu\to\jpsi\rhop$ backgrounds have a similar distribution in the \jpsi\piz mass.
In the default fit model, the yield of the latter is free to vary, while the former yield is fixed to the value determined in another fit to the diphoton mass (see Sec.~\ref{sec:fitmodel_pi0}).
A model where the \KS yield is free to vary and the $\Bu\to\jpsi\rhop$ yield fixed to its expected value (as determined in Sec.~\ref{sec:checks_bkg}) gives a small 0.4\% change in the results from the default model.
Similarly, fixing the yield of \Bc decays has a 0.7\% effect on the signal yield.

In the default normalisation fit, random photon background contributions are constrained using calibrated simulation samples, while a data-driven approach is adopted in the signal fit.
The normalisation yield, however, is robust against the choice of model for these backgrounds with a small variation of 0.4\% when the signal model is used, which is at the level of the associated systematic uncertainty of 0.3\%.

By default, no action is taken to remove multiple candidates from the signal and normalisation data samples to which they contribute at the level of one and five parts per thousand, respectively.
The bias in the yields when randomly removing multiple candidates from the samples is indeed found to be negligible.

\section{Results}
\label{sec:results}

The obtained branching fraction ratio between $\Bz\to\jpsi\piz$ and $\Bu\to\jpsi\Kstarp$ decays is

\begin{equation*}
    \mathcal{R} = (1.153 \pm  0.053~(\textrm{stat.}) \pm 0.048~(\textrm{syst.})) \times 10^{-2}.
\label{eq:result2}
\end{equation*}

\noindent
The current world-average for the normalisation branching fraction of $(1.43\pm0.08)\times10^{-3}$ \cite{PDG2022} includes seven measurements performed using \Kstarp candidates with a reconstructed mass within a $100\mevcc$ mass window or smaller of the known \Kstarp mass \cite{BaBar:2007esv,BaBar:2004htr,Belle:2002otd,CLEO:1997ilq,CDF:1995izg,CLEO:1991roe,ARGUS:1990jet}.
The \belle analysis \cite{Belle:2002otd} is the only to measure the yield for resonant decays but estimates the contributions from nonresonant decays at the level of 5.2\% in the \Kstar mass region.
Including the nonresonant contributions, the uncertainty-weighted average of the normalisation branching fraction increases to

\begin{equation*}
    \BF_{\Bu\to\jpsi\Kstarp} = (1.449 \pm 0.083) \times 10^{-3}.
\label{eq:result3}
\end{equation*}

\noindent
Combined with the world average $\BF_{\Bz\to\jpsi\piz} = (1.66 \pm 0.10) \times 10^{-5}$, the expected ratio $\mathcal{R}$ of $(1.15 \pm 0.10)\times 10^{-2}$ agrees with the result of this analysis, the latter being more precise.

\begin{figure}[htb!]
\begin{center}
\scalebox{0.5}{\input{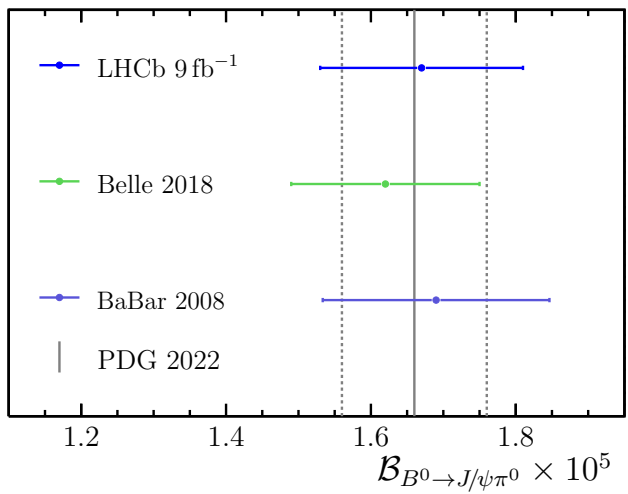}}
\caption{\small Branching fraction for $\Bz\to\jpsi\piz$ decays as measured in this analysis and by the \babar and \belle collaborations. The world average and its uncertainty (PDG 2022) are indicated by the solid and dashed lines, respectively, representing a combination of the \belle, \babar and \cleo measurements \cite{Belle:2018nxw,BaBar:2008kfx,PhysRevD.62.051101}. The latter measurement is affected by a much larger error and is not shown.}
\label{fig:bf_result}
\end{center}
\end{figure}

Finally, using this new average for the normalisation branching fraction, the obtained branching fraction for signal decays is

\begin{equation*}
    \BF_{\Bz\to\jpsi\piz} = (1.670 \pm 0.077~(\textrm{stat.}) \pm 0.069~(\textrm{syst.}) \pm 0.095~(\textrm{ext.})) \times 10^{-5},
\end{equation*}

\noindent
where the last uncertainty reflects the precision on the branching fraction of the normalisation mode. As shown in Fig.~\ref{fig:bf_result}, this result is in agreement with previous measurements.

\section{Summary}
\label{sec:summary}

The measurement of the branching fraction of $\Bz\to\jpsi\piz$ decays using $\Bu\to\jpsi\Kstarp$ decays as normalisation channel is presented. The measured ratio of branching fractions is compatible with the value determined from the average branching fractions of the individual decay modes and is more precise.
Using a slightly increased average for the normalisation mode that accounts for the presence of nonresonant decays in the analysed $\Bu\to\jpsi\Kstarp$ sample, the branching fraction obtained for $\Bz\to\jpsi\piz$ decays is $(1.67 \pm 0.14)\times10^{-5}$, compatible with the current world-average of $(1.66 \pm 0.10) \times 10^{-5}$.
This result achieves a similar precision as the most precise single measurement of $(1.62 \pm 0.13) \times 10^{-5}$ published by the \belle collaboration \cite{Belle:2018nxw}. This analysis also paves the way for future \CP-violation measurements with these decays.

\clearpage


\section*{Acknowledgements}
%
%
\noindent We express our gratitude to our colleagues in the CERN
accelerator departments for the excellent performance of the LHC. We
thank the technical and administrative staff at the LHCb
institutes.
We acknowledge support from CERN and from the national agencies:
CAPES, CNPq, FAPERJ and FINEP (Brazil); 
MOST and NSFC (China); 
CNRS/IN2P3 (France); 
BMBF, DFG and MPG (Germany); 
INFN (Italy); 
NWO (Netherlands); 
MNiSW and NCN (Poland); 
MCID/IFA (Romania); 
MICINN (Spain); 
SNSF and SER (Switzerland); 
NASU (Ukraine); 
STFC (United Kingdom); 
DOE NP and NSF (USA).
We acknowledge the computing resources that are provided by CERN, IN2P3
(France), KIT and DESY (Germany), INFN (Italy), SURF (Netherlands),
PIC (Spain), GridPP (United Kingdom), 
CSCS (Switzerland), IFIN-HH (Romania), CBPF (Brazil),
and Polish WLCG (Poland).
We are indebted to the communities behind the multiple open-source
software packages on which we depend.
Individual groups or members have received support from
ARC and ARDC (Australia);
Key Research Program of Frontier Sciences of CAS, CAS PIFI, CAS CCEPP, 
Fundamental Research Funds for the Central Universities, 
and Sci. \& Tech. Program of Guangzhou (China);
Minciencias (Colombia);
EPLANET, Marie Sk\l{}odowska-Curie Actions, ERC and NextGenerationEU (European Union);
A*MIDEX, ANR, IPhU and Labex P2IO, and R\'{e}gion Auvergne-Rh\^{o}ne-Alpes (France);
AvH Foundation (Germany);
ICSC (Italy); 
GVA, XuntaGal, GENCAT, Inditex, InTalent and Prog.~Atracci\'on Talento, CM (Spain);
SRC (Sweden);
the Leverhulme Trust, the Royal Society
 and UKRI (United Kingdom).

\addcontentsline{toc}{section}{References}
\bibliographystyle{LHCb}
\bibliography{main,standard,LHCb-PAPER,LHCb-CONF,LHCb-DP,LHCb-TDR}

\newpage
\centerline
{\large\bf LHCb collaboration}
\begin
{flushleft}
\small
R.~Aaij$^{35}$\lhcborcid{0000-0003-0533-1952},
A.S.W.~Abdelmotteleb$^{54}$\lhcborcid{0000-0001-7905-0542},
C.~Abellan~Beteta$^{48}$,
F.~Abudin{\'e}n$^{54}$\lhcborcid{0000-0002-6737-3528},
T.~Ackernley$^{58}$\lhcborcid{0000-0002-5951-3498},
J. A. ~Adams$^{66}$\lhcborcid{0009-0003-9175-689X},
A. A. ~Adefisoye$^{66}$\lhcborcid{0000-0003-2448-1550},
B.~Adeva$^{44}$\lhcborcid{0000-0001-9756-3712},
M.~Adinolfi$^{52}$\lhcborcid{0000-0002-1326-1264},
P.~Adlarson$^{79}$\lhcborcid{0000-0001-6280-3851}
% ...

{\footnotesize \it

% ...
$^{35}$Nikhef National Institute for Subatomic Physics, Amsterdam, Netherlands\\
% ...
$^{44}$Instituto Galego de F{\'\i}sica de Altas Enerx{\'\i}as (IGFAE), Universidade de Santiago de Compostela, Santiago de Compostela, Spain\\
% ...
$^{48}$Physik-Institut, Universit{\"a}t Z{\"u}rich, Z{\"u}rich, Switzerland\\
% ...
$^{52}$H.H. Wills Physics Laboratory, University of Bristol, Bristol, United Kingdom\\
% ...
$^{54}$Department of Physics, University of Warwick, Coventry, United Kingdom\\
% ...
$^{57}$School of Physics and Astronomy, University of Glasgow, Glasgow, United Kingdom\\
$^{58}$Oliver Lodge Laboratory, University of Liverpool, Liverpool, United Kingdom\\
% ...
$^{66}$Syracuse University, Syracuse, NY, United States\\
% ...
$^{79}$Department of Physics and Astronomy, Uppsala University, Uppsala, Sweden, associated to $^{57}$\\
% ...
}
\end{flushleft}

\end{document}